\documentclass[journal]{IEEEtran}

\usepackage{graphicx}
\usepackage{xcolor}
\usepackage{layouts}
\usepackage{float}
\usepackage{caption}
\usepackage{subcaption}
\usepackage{multirow}
\usepackage[ruled,vlined,linesnumbered]{algorithm2e}
\usepackage{ragged2e}
\usepackage{amsmath}
\usepackage{tcolorbox}
\tcbset{center title}
\usepackage{hyperref}
\hypersetup{
    colorlinks = true, %Colour links instead of ugly boxes
    urlcolor = black, %Colour for external hyperlinks
    linkcolor = blue, %Colour of internal links
    citecolor = blue %Colour of citations
}

\begin{document}

\begin{center}
\begin{tcolorbox}[width=\textwidth, halign=center, valign=center, title={NOTICE}]
This work has been submitted to the IEEE for possible publication. Copyright may be transferred without notice, after which this version may no longer be accessible.
\end{tcolorbox}
\end{center}

\title{Data Criticality in Multi-Threaded Applications:\\An Insight for Many-Core Systems}
\author{Abhijit~Das,~\IEEEmembership{Student~Member,~IEEE,}
        John~Jose,~\IEEEmembership{Member,~IEEE,}
        and~Prabhat~Mishra,~\IEEEmembership{Fellow,~IEEE}% <-this % stops a space
\thanks{Manuscript received Month 00, 0000; revised Month 00, 0000; accepted Month 00, 0000. Date of publication Month 00, 0000; date of current version Month 00, 0000. \textit{(Corresponding author: Abhijit Das.)}}
\thanks{A. Das and J. Jose are with the Department of Computer Science and Engineering, Indian Institute of Technology Guwahati, Assam 781039, India (e-mail: abhijit.das@iitg.ac.in; johnjose@iitg.ac.in).}% <-this % stops a space
\thanks{P. Mishra is with the Department of Computer and Information Science and Engineering, University of Florida, Gainesville, FL 32611, USA.}}

% The paper headers
% \markboth{IEEE TRANSACTIONS ON VERY LARGE SCALE INTEGRATION (VLSI) SYSTEMS, VOL.~00, NO.~00, MONTH~0000}{}

\maketitle

\begin{abstract}
Multi-threaded applications are capable of exploiting the full potential of many-core systems. However, Network-on-Chip (NoC) based inter-core communication in many-core systems is responsible for 60-75\% of the miss latency experienced by multi-threaded applications. Delay in the arrival of critical data at the requesting core severely hampers performance. This brief presents some interesting insights about how critical data is requested from the memory by multi-threaded applications. Then it investigates the cause of delay in NoC and how it affects the performance. Finally, this brief shows how NoC-aware memory access optimisations can significantly improve performance. Our experimental evaluation considers {\it early restart}  memory access optimisation and demonstrates that by exploiting NoC resources, critical data can be prioritised to reduce miss penalty by 10--12\% and improve system performance by 7--11\%.
\end{abstract}

% Note that keywords are not normally used for peer review papers.

\section{Introduction}
\IEEEPARstart{A}{pplications} running in any computing device can be broadly classified as either multi-programmed or multi-threaded. Many-core systems have made way for applications with massive processing requirements; something which was not possible earlier. The processing power of many-core systems come from a collection of relatively simpler processing cores, unlike a single powerful core in uni-core systems. Hence, to exploit the full potential of many-core systems, applications need to be parallel, thus multi-threaded. Modern many-core systems employ Network-on-Chip (NoC) based inter-core communication, and it is reported that NoC is responsible for 60--75\% of the miss latency in multi-threaded applications~\cite{sanchez2010analysis}. Delay in arrival of the critical data at the requesting core hampers performance of such applications. Hence, it is very important to get an insight into how multi-threaded applications request critical data and how NoC contributes to the miss latency (miss penalty) of such applications.

\begin{figure}[t]
	\centering
	\includegraphics[width=\columnwidth]{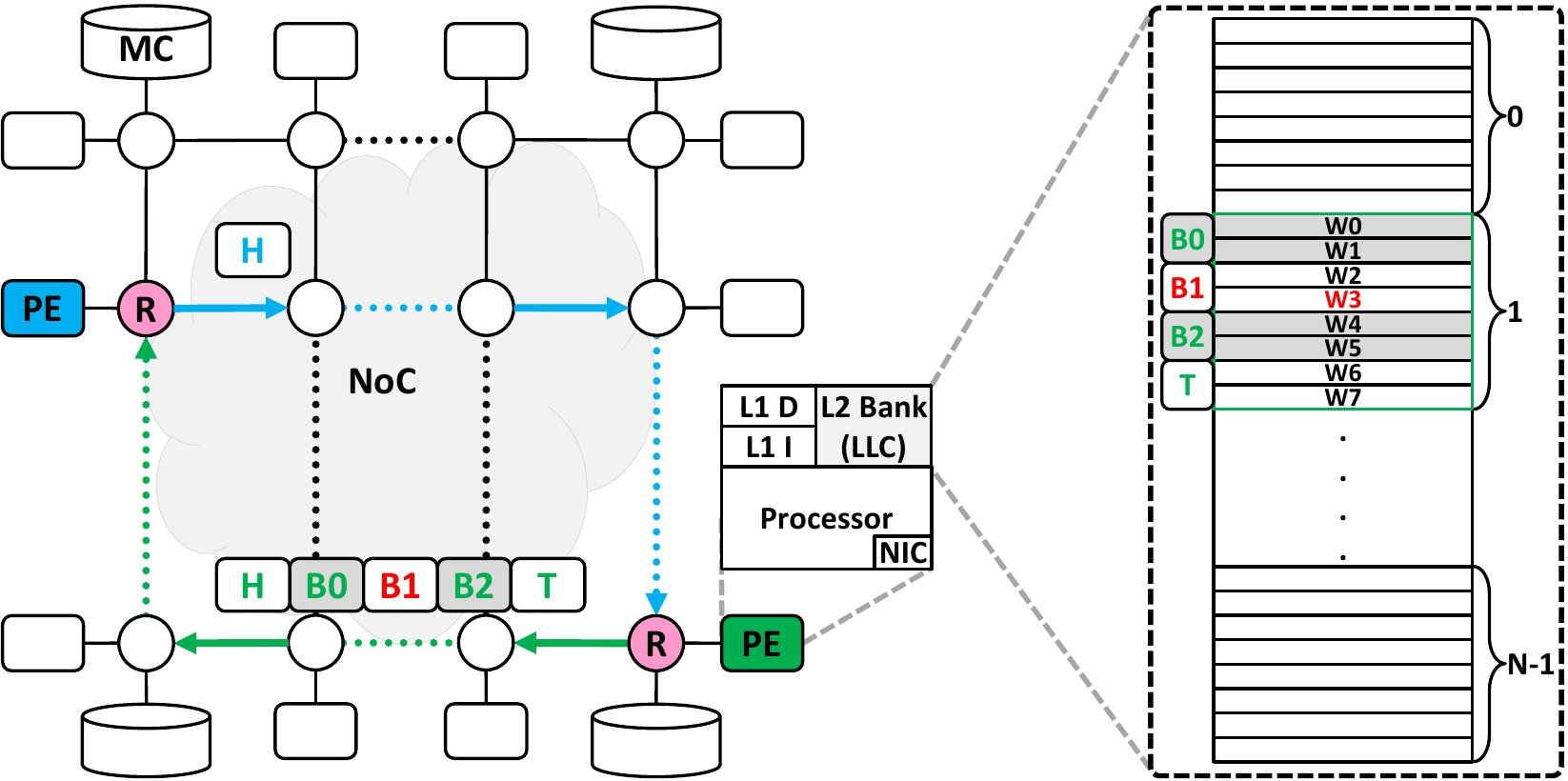}
	\caption{Conceptual view of an NoC based many-core system where, PE: Processing Element, R: Router, LLC: Last Level Cache and MC: Memory Controller. Due to limited on-chip data transfer bandwidth, a packet in NoC is divided into multiple smaller units called flits. A request packet consists of a single head flit (H), whereas a reply packet consists of a head flit followed by multiple body flits and ended by a tail flit (H, B0, B1, B2, T). Flits of a particular packet always travel in order. A critical word (W3) and the flit carrying that critical word (B1) are shown in \textit{red} for reference. The request path  is shown in blue, whereas the reply path is shown in green.}
	\label{fig:tcmp}
	\vspace{-0.6cm}
\end{figure}

While running an application, a core usually requests for a single word from memory, called \textit{critical word}~\cite{hennessy2011computer}. The critical word is first searched in the cache memory and returned to the core if found. However, if the word is not found in the cache, it is requested from the next level of memory. The smallest unit of data transfer between different levels of memory is in blocks, where a block consists of multiple words. So, even though the core requests a single word, a complete block (containing the critical word) is brought from the next level of memory in the form of a packet through the underlying NoC. Nevertheless, transfer bandwidth in NoC is limited to channel width called \textit{flit}. A data block (packet) is divided into multiple flits (flit $<<$ block) and sent in sequence, as shown in Figure~\ref{fig:tcmp}. The critical word can be in any of the incoming flits and hence would accordingly impact the performance. On their way, the incoming flits experience indefinite router delays due to congestion, which impact their arrival on the requesting core, which again hampers the performance.

The purpose of this brief is to share some interesting insights about running multi-threaded applications in many-core systems. It specifically presents the pattern of critical words requested from the memory. Then it describes the pattern of delays experienced by the flits while travelling from source to their destination. The brief finally shows how using these insights can significantly improve the performance of existing optimisations on miss penalty reduction. Specifically, this brief makes the following major contributions:

\begin{figure*}
     \centering
     \begin{subfigure}[c]{0.496\textwidth}
         \centering
         \includegraphics[width=\textwidth, height=3cm]{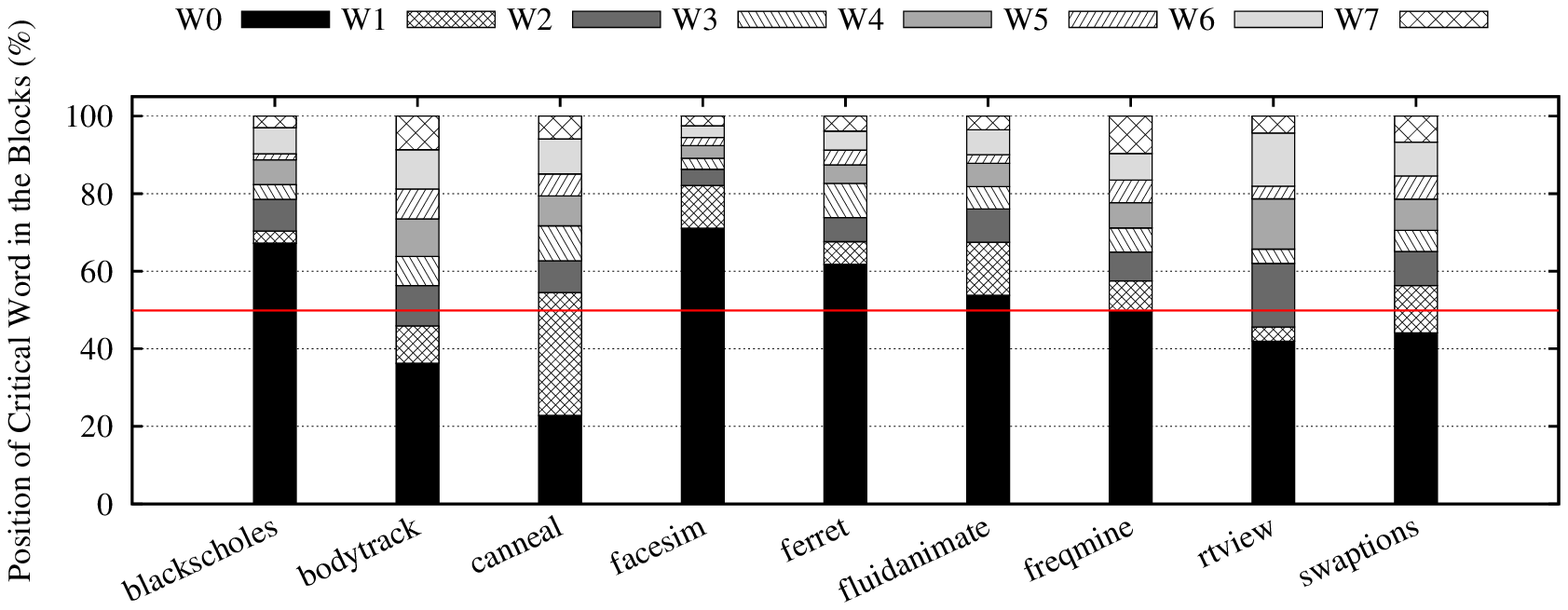}
         \caption{PARSEC 3.0}
         \label{fig:cwp-p}
     \end{subfigure}
     %\hfill
     \begin{subfigure}[c]{0.496\textwidth}
         \centering
         \includegraphics[width=\textwidth, height=3cm]{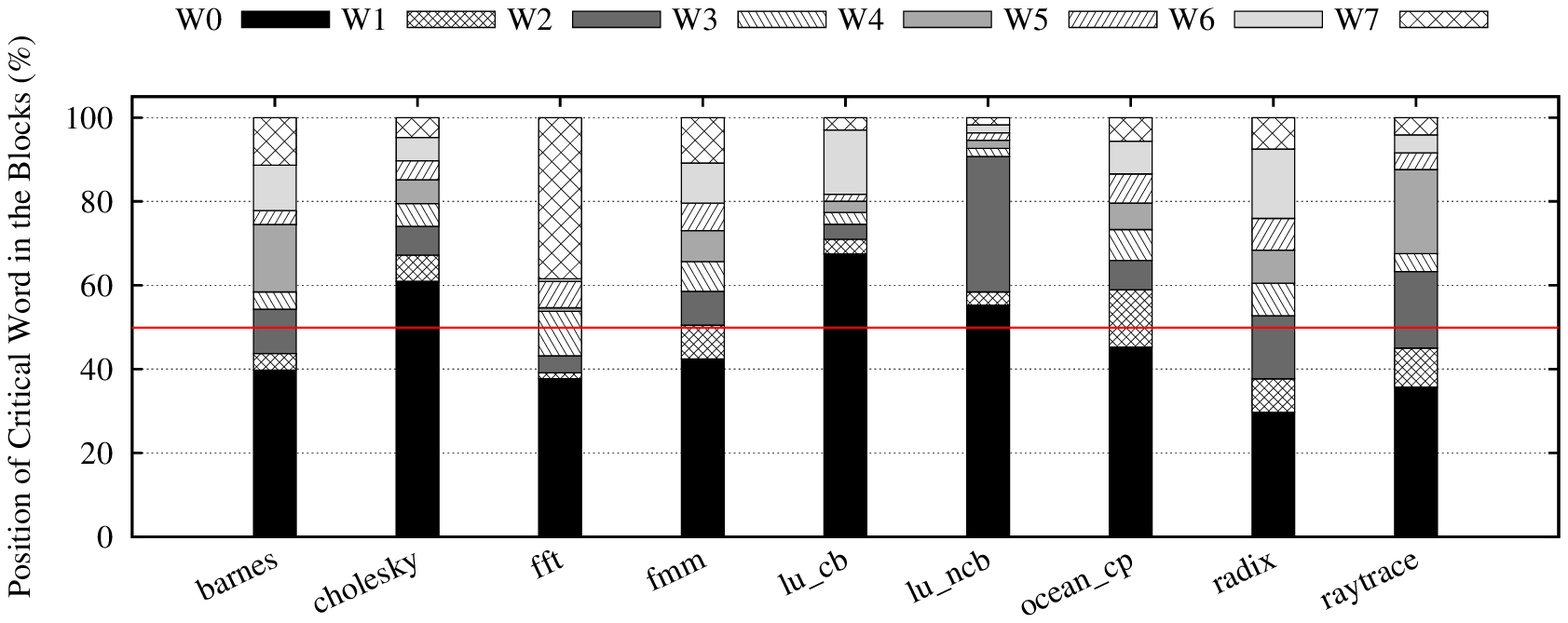}
         \caption{SPLASH-2x}
         \label{fig:cwp-s}
     \end{subfigure}
     \caption{Position of critical word in the requested blocks}
     \label{fig:cwp}
     \vspace{-0.6cm}
\end{figure*}

\begin{enumerate}
    \item In a first, it presents the position of critical word within the requested block and the corresponding flits for PARSEC 3.0~\cite{bienia2008parsec} and SPLASH-2x~\cite{woo1995splash} benchmarks.
    \item It describes the difference in arrival times of the first and last flit of an incoming data block and how it impacts the performance of the underlying applications. 
    \item It proposes a modified version of the popular {\it early restart}  optimisation~\cite{hennessy2011computer} by considering the insights from (1) and (2), which performs better than the original.
\end{enumerate}

\section{Data Criticality in Applications}
In a multi-threaded application, multiple tasks (threads) may run concurrently and independently by using the shared resources. This is the reason why multi-threaded applications utilise the resources of many-core systems better. Current many-core systems like Intel Xeon Phi Processor
(2016)~\cite{sodani2016knights}, Princeton Piton Processor (2015)~\cite{balkind2016openpiton}, MIT Scorpio
Processor (2014)~\cite{daya2014scorpio}, and others have private L1 caches and a shared L2 cache. In most of these processors, the L2 cache is divided into multiple banks and distributed across all the cores, as shown in Figure~\ref{fig:tcmp}. When the critical word is not found in the L1 cache (miss), the core requests the word from the corresponding L2 cache bank. The entire block containing the critical word is transferred from L2 to L1 cache (refer Figure~\ref{fig:tcmp}). In a conventional system, even though the core requires only the critical word to resume its execution, it is made to wait till the arrival of the entire block. We know that a block contains multiple words, so imagine a scenario where the very first word of the block is the critical word. The core could resume its execution after the arrival of the first word, but instead, it needs to wait till the last word of the block. Hence we are motivated to profile state-of-the-art multi-threaded applications to know the pattern in which critical words are spread in their corresponding block. This insight will help us to understand how popular memory access optimisations like \textit{Early Restart} and \textit{Critical Word First}~\cite{hennessy2011computer} can benefit the corresponding applications (explained in Section~\ref{proposal}).

We profile PARSEC 3.0~\cite{bienia2008parsec} and SPLASH-2x~\cite{woo1995splash}, the two most popular suite of multi-threaded benchmarks. We model an equivalent implementation of Intel Xeon Phi Processor 7235~\cite{intel7235}, one of the latest many-core systems, on gem5 simulator~\cite{binkert2011gem5}. Our system configuration is given in Table~\ref{tab:config} for reference. We profile those data requests for whom the critical word was not found in the L1 cache. These are the requests that travel through the underlying NoC to reach L2 cache bank and get data, thus suffers NoC related delay~\cite{sanchez2010analysis}. To the best of our knowledge, this profiling is a first of its kind for any multi-threaded applications. Figure~\ref{fig:cwp} shows the average position of critical word in the corresponding blocks requested from L2 cache. To understand with an example, during the entire run of \textit{blackcholes} benchmark from PARSEC 3.0 suite, for 67.20\% of the time, the first word is the critical word, for 3.17\% of the time, the second word is the critical word, for 8.17\% of the time, the third word is the critical word and so on. There is a very interesting trend: {\it the first word is the critical word for most of the requested blocks}. This pattern is observed for the majority of the benchmarks of both PARSEC 3.0 and SPLASH-2x suites even though they are from diverse domains like computer vision, media processing, computational finance, enterprise servers, animation physics, high-performance computing, etc. However, the trend is unusual but not unreasonable, as the existing literature has proof of critical word regularity~\cite{gieske2008critical}\cite{chatterjee2012leveraging}. Literature states that it is reasonable to expect that data in a given region may be accessed in similar order on multiple occasions.

\begin{table}[t]
    \renewcommand{\arraystretch}{1.3}
	\caption{System configuration}
	\label{tab:config}
	\begin{center}
	    \resizebox{\columnwidth}{!}
		{
		    \begin{tabular}{l|lc|c}
		        \hline
				\bf {Processor} & 64 OoO x86 cores \\
				\hline
				\bf {L1 Cache} & 32KB, 8-way, private, split \\
				\hline
				\bf {L2 Cache (LLC)} & 512KB$\times$64 cores, 16-way, shared \\
				\hline
				\bf {Memory Bank} & 4; one located at each corner \\
				\hline
				\bf {NoC} & 8$\times$8 2D mesh, 3-VCs/port, 128-bit channel \\
				\hline
				\bf {Routing} & 2-stage routers, X-Y dimension-order routing \\
				\hline
				\bf {Packets} & 1-flit for control packets, 5-flit for data packets \\
				\hline
				\bf {Word/Flit/Block} & 64-bit/128-bit/68B; 2-words/flit, 8-words/block \\
				\hline
				\bf {Benchmarks} & PARSEC 3.0 and SPLASH-2x (multi-threaded) \\
				\hline
			\end{tabular}
		}
	\end{center}
	\vspace{-0.8cm}
\end{table}

While explaining the individual memory access patterns for each of the benchmarks is beyond the scope of this brief, we give some common characteristics that justify the pattern on the location of a critical word. Benchmarks that traverse through data arrays exhibit critical words near the beginning of the data block, most often to word 0. Also, the benchmarks having strided access with the smallest stride length of 0 have word 0 as the critical word. On the other hand, there are also benchmarks like \textit{canneal} and \textit{radix}, where the critical word is somewhat uniformly distributed. Benchmarks whose memory accesses are generated due to pointer chasing exhibits better distribution of the critical word. Based on these observations, specific memory access optimisation can be implemented for a class of applications exhibiting a specific behaviour to reduce the miss penalty of the critical word and improve performance.

\begin{table}[t]
    \renewcommand{\arraystretch}{1.3}
    \caption{Position of critical word in the corresponding flits. Note that head flit (H) does not carry any words of the data block and hence its corresponding column is left empty.}
	\label{tab:benchchar}
    \begin{center}
        \resizebox{\columnwidth}{!}
		{
            \begin{tabular}{l|l|l|l|l|l|l|l}
            \hline
            \multirow{2}{*}{\textbf{\#}} & \multirow{2}{*}{\textbf{Suite}} & \multirow{2}{*}{\textbf{Benchmark}} & \multicolumn{5}{l}{\textbf{Flits of an incoming requested block}} \\ \cline{4-8}
                       &                 &               & \textbf{H} & \textbf{B0} & \textbf{B1} & \textbf{B2} & \textbf{T} \\ \hline
            1          & PARSEC 3.0      & blackscholes  &            & 70.37	    & 11.94       & 7.95	    & 9.74       \\ \hline
            2          & PARSEC 3.0      & bodytrack     &            & 45.90	    & 17.91	      & 17.38	    & 18.81      \\ \hline
            3          & PARSEC 3.0      & canneal       &            & 54.49	    & 17.20	      & 13.38	    & 14.93      \\ \hline
            4          & PARSEC 3.0      & facesim       &            & 82.06	    & 7.06	      & 5.40	    & 5.48       \\ \hline
            5          & PARSEC 3.0      & ferret        &            & 67.60	    & 15.04	      & 8.56    	& 8.80       \\ \hline
            6          & PARSEC 3.0      & fluidanimate  &            & 67.47	    & 14.34	      & 8.23	    & 9.96       \\ \hline
            7          & PARSEC 3.0      & freqmine      &            & 57.50	    & 13.67	      & 12.32   	& 16.51      \\ \hline
            8          & PARSEC 3.0      & rtview        &            & 45.59	    & 20.08	      & 16.26   	& 18.07      \\ \hline
            9          & PARSEC 3.0      & swaptions     &            & 56.33	    & 14.21	      & 14.04   	& 15.42      \\ \hline
            10         & SPLASH-2X       & barnes        &            & 43.70	    & 14.69	      & 19.4	    & 22.21      \\ \hline
            11         & SPLASH-2X       & cholesky      &            & 67.19	    & 12.27	      & 10.26   	& 10.28      \\ \hline
            12         & SPLASH-2X       & fft           &            & 39.10	    & 14.69	      & 7.14    	& 39.07      \\ \hline
            13         & SPLASH-2X       & fmm           &            & 50.46	    & 15.19	      & 13.94   	& 20.41      \\ \hline
            14         & SPLASH-2X       & lu\_cb        &            & 70.99	    & 6.37	      & 4.32    	& 18.32      \\ \hline
            15         & SPLASH-2X       & lu\_ncb       &            & 58.41	    & 34.25	      & 3.70	    & 3.64       \\ \hline
            16         & SPLASH-2X       & ocean\_cp     &            & 58.96	    & 14.35	      & 13.27   	& 13.42      \\ \hline
            17         & SPLASH-2X       & radix         &            & 37.68	    & 22.81	      & 15.47	    & 24.04      \\ \hline
            18         & SPLASH-2X       & raytrace      &            & 45.03	    & 22.51	      & 24.03   	& 8.43       \\ \hline
            \end{tabular}
		}
	\end{center}
	\vspace{-0.8cm}
\end{table}

\section{Criticality Aware Many-Core Systems}\label{proposal}
The most popular memory access optimisations to reduce miss penalty of the critical word are \textit{early restart} and \textit{critical word first}~\cite{hennessy2011computer}. In {\it early restart}, as soon as the critical word is received at the L1 cache, it is forwarded to the processor to resume its execution without waiting for the entire block. In {\it critical word first}, the critical word is forwarded out-of-order by the L2 cache to be received as the first word in the L1 cache to resume processor execution at the earliest. In NoC based many-core systems, everything travels as packets, including data blocks. Due to limited on-chip channel width, packets are further divided into multiple flits. A head flit (H) carries the packet (message) header containing the routing information and does not carry any part of the data block. Multiple body flits (Bi), ended by a tail flit (T) carries the data block from source to the destination. Hence, a data block (of 8-words, refer Table~\ref{tab:config}) is transferred as a sequence of head flit, followed by three body flits (of 2-words each) and a tail flit (of 2-words) (H, B0, B1, B2, T). Table~\ref{tab:benchchar} presents the percentage of critical words that fall on different flits of a data block.

To understand the observation, when the requested data block in \textit{blackscholes} benchmark of PARSEC 3.0 suite is transferred through flits, 70.37\% of the time, critical words are in flit B0, 11.94\% in flit B1, 7.95\% in flit B2 and 9.74\% in the tail flit T. It can be seen that the first body flit (B0) contains the critical words most of the time. It was evident from the fact that most of the critical words are the first word of a data block (refer Figure~\ref{fig:cwp}) and B0 carries the first two words of the block. Hence, {\it early restart}  optimisation can be very effective in these kinds of applications. {\it Critical word first}  optimisation involves sending the critical word in the first flit by allowing out of order travel. Since by their very nature, the majority of the applications have their critical word in the first flit itself, {\it critical word first}  might not be required. Avoiding {\it critical word first}  also brings in the advantage of avoiding the complexity of sending the word out of order and then reordering the words at their destination.

Both {\it early restart}  and {\it critical word first}  optimisations are oblivious of the underlying on-chip communication infrastructure. These optimisations were introduced in the era of bus based on-chip communication, where a data block is transferred as a continuous stream of words. Hence, the data block is transferred within a fixed time, and the core could resume execution at the earliest as per the corresponding optimisation. However, modern many-core systems use NoC based on-chip communication where the data block is transferred as multiple flits in a discrete fashion. On their way to the destination, flits experience indefinite router delay due to on-chip congestion. As a result, the effectiveness of {\it early restart}  and {\it critical word first}  reduces in many-core systems. 

\begin{figure}[t]
	\centering
	\includegraphics[width=\columnwidth, height=3cm]{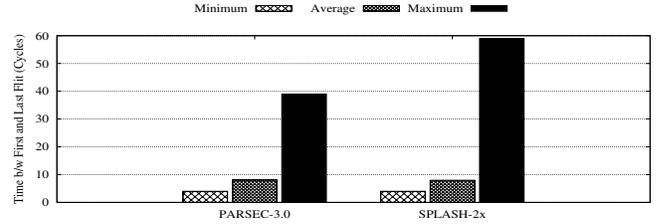}
	\caption{Reply Difference Time}
	\label{fig:rdt}
	\vspace{-0.6cm}
\end{figure}

To understand about the delay experienced by incoming flits, we conduct an experimental analysis for all the PARSEC 3.0 and SPLASH-2x benchmarks. We consider a metric from the literature called, \textit{Reply Difference Time (RDT)}~\cite{das2018critical} which calculates the difference between the arrival of the first flit and last flit of a data block in the requesting core. Figure~\ref{fig:rdt} shows the average, minimum and maximum time difference between the arrival of the first and last flit of the data block. The minimum RDT remains 4 for both PARSEC 3.0 and SPLASH-2x benchmarks. It implies that all the flits reach back to back without any delay (ideal case). However, the maximum RDT is surprisingly high (during congestion): 39 for PARSEC 3.0 and 59 for SPLASH-2x benchmarks. Even the average RDT is 8.01 and 7.98, more than the minimum RDT meaning, flits are generally getting delayed. If memory access optimisations are made aware of the pros and cons of the underlying on-chip communication infrastructure, they may yield even better benefits. The purpose of this brief is not to provide NoC-aware implementations of all the existing critical word based memory access optimisations. Instead, we take just one of the most popular memory access optimisations, {\it early restart}, and demonstrate that an NoC-aware {\it early restart} is effective in significantly improving the overall system performance.

\section{NoC-Aware Early Restart Optimisation}
This section proposes an NoC-aware {\it early restart} optimisation in many-core systems based on the observations from Figure~\ref{fig:cwp}, Table~\ref{tab:benchchar} and Figure~\ref{fig:rdt}. For the ease of illustration, we call the original {\it early restart}  optimisation as \textbf{ER} and our proposed version of NoC aware {\it early restart}  as \textbf{ER-NoC}.

\subsection{Critical Flit Identification}
\begin{figure}[t]
	\centering
	\includegraphics[width=\columnwidth]{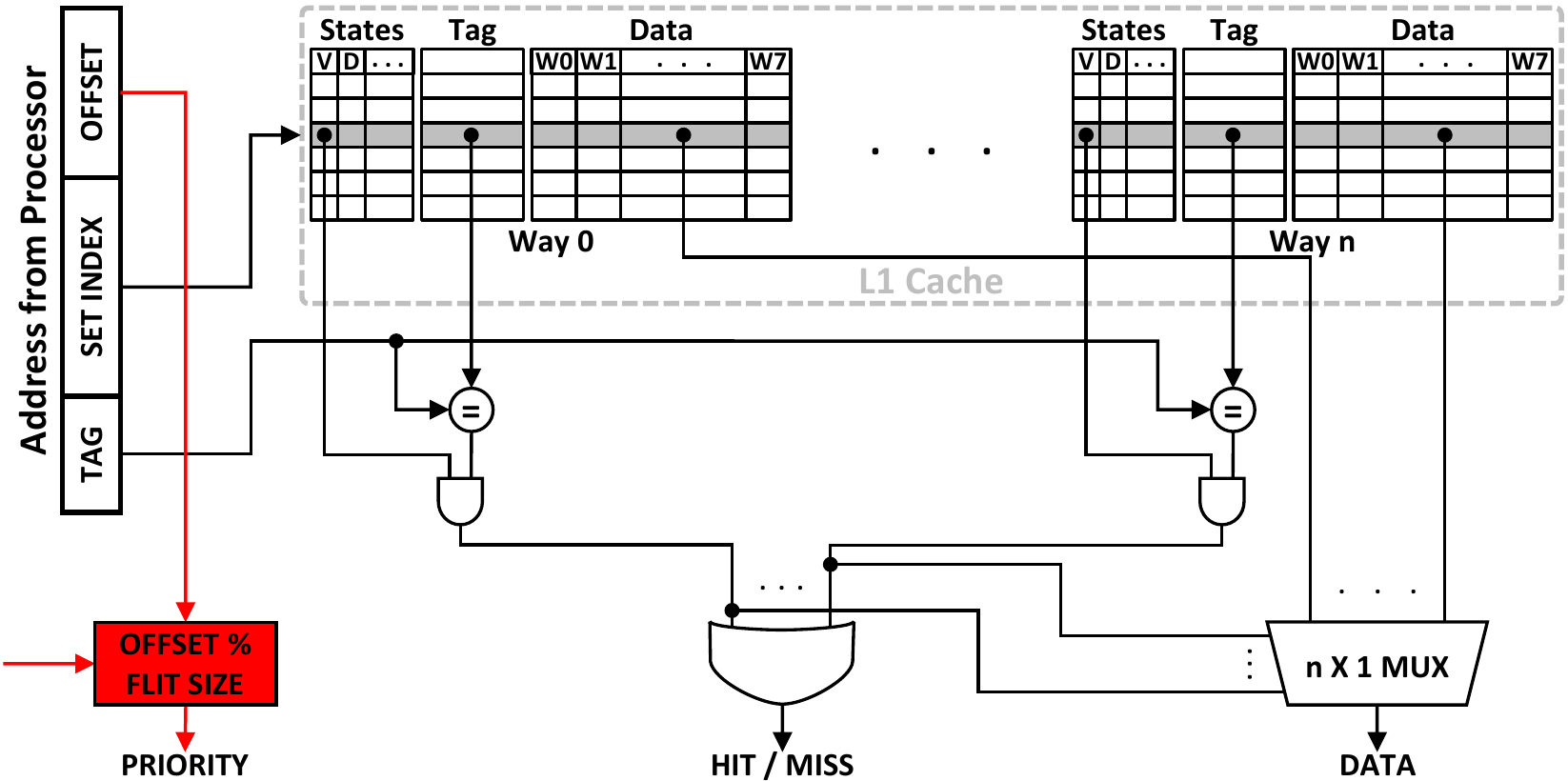}
	\caption{Modified L1 cache controller}
	\label{fig:l1-ctlr}
	\vspace{-0.6cm}
\end{figure}

When the core (processor) needs a critical word from the private L1 cache, it sends a request to the L1 cache controller (L1 CTLR) giving the address of the corresponding data block which contains the word. L1 CTLR maps into the corresponding set using the set index bits of the requested address, as shown in Figure~\ref{fig:l1-ctlr}. After the set is identified, L1 CTLR checks the tag bits for hit/miss detection. The corresponding data lookup is also done in parallel to reduce memory access latency. If the tag checker returns true (cache hit), the requested data block is present in the L1 cache. L1 CTLR identifies the critical word using the offset bits of the given address and sends it to the processor. However, if the tag checker returns false (cache miss), the data block needs to be brought from the corresponding L2 cache bank through the underlying NoC. While the block is being fetched from the L2 cache bank, ER optimisation uses the offset bits to check for the arrival of the critical word. As soon as the critical word is fetched at L1 cache, it is sent to the processor to resume its execution, without waiting for the arrival of the entire block. While the processor resumes execution, the remaining words of the block are fetched in the background. Since data transfer is still performed at block-level granularity, the underlying cache coherence remains unaffected.

We define \textit{critical flit} as the flit carrying the critical word of the block through the underlying NoC. The proposed ER-NoC optimisation adds a tiny module at L1 CTLR to identify the critical flit, as shown in red in Figure~\ref{fig:l1-ctlr}. This module takes as input the offset bits and flit size to return a 2-bit value (00/01/10/11) called \textit{critical flit identifier (CFI)}. If the CFI value is 00, it means that the critical word will be transferred in the body flit B0, if 01 then B1, if 10 then B2 and if 11 then in the tail flit T. The proposed module added in L1 CTLR to get CFI runs in parallel with tag checker and data lookup modules and hence does not incur any additional delay in memory access latency. If the tag checker returns true, the CFI value is ignored as the requested block is already in L1 cache and there will be no flit transfer. However, if the tag checker returns false, the CFI value is added to the message/packet header when the block request is sent to the L2 cache bank.

\subsection{Critical Flit Prioritisation}
\begin{figure}[t]
	\centering
	\includegraphics[width=\columnwidth]{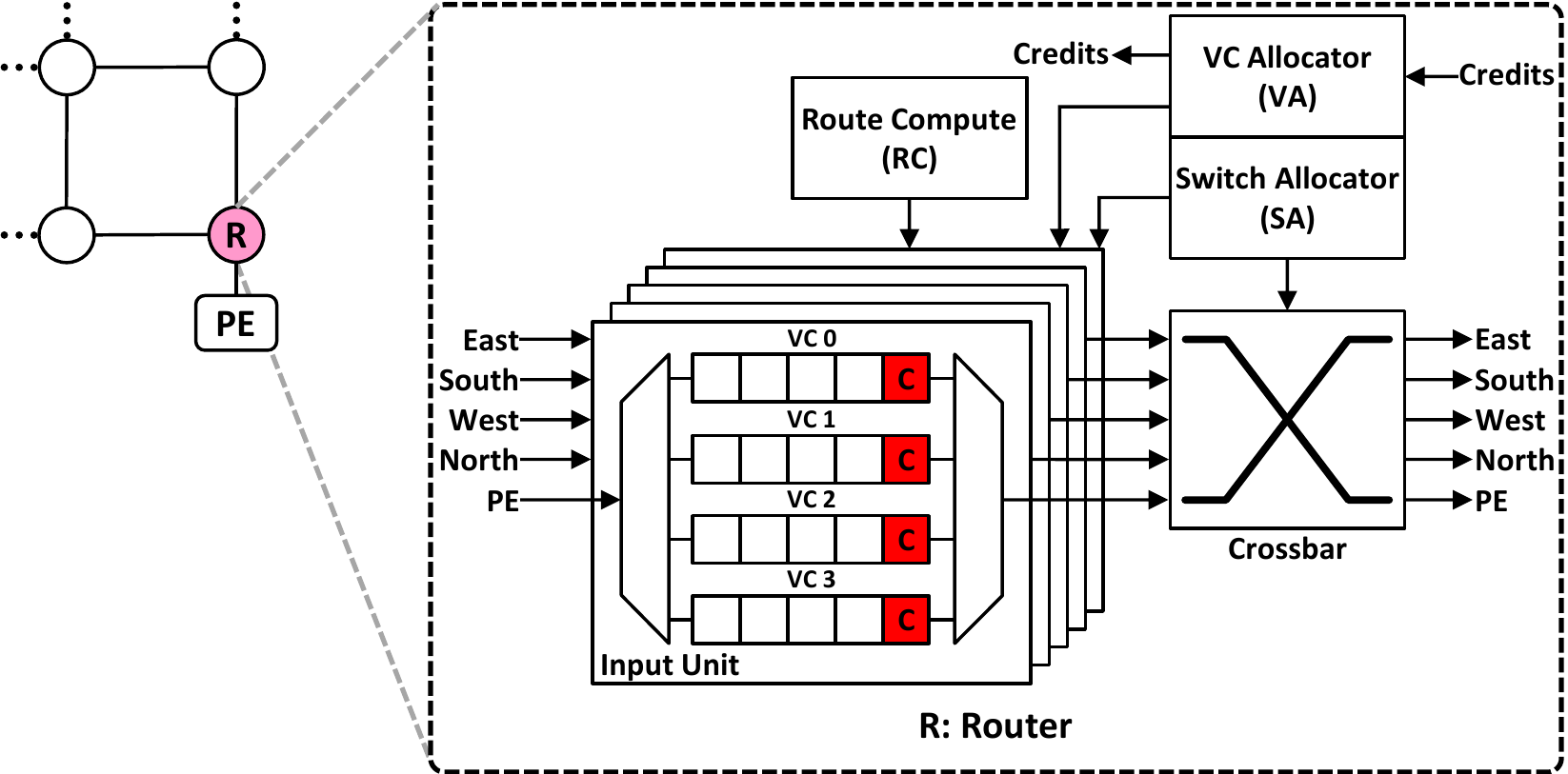}
	\caption{Modified router microarchitecture}
	\label{fig:router}
	\vspace{-0.6cm}
\end{figure}

Upon receiving the request, the L2 cache bank controller (L2 CTLR) replies with the data block as multiple flits through the underlying NoC. For the proposed ER-NoC optimisation, L2 CTLR puts the corresponding CFI value back into the packet header (H) of the reply. When the head flit traverse through the NoC routers towards its destination, the CFI values are stored in a counter \textit{C} in each router, as shown in Figure~\ref{fig:router}. ER-NoC modifies the traditional round-robin based priority policy to use the CFI counter for priority during routing and arbitration. The motive behind this move is to reduce the router delay for critical flits so that they reach the destination at the earliest. However, with priority, there is always a risk of starvation for lower priority flits. To minimise such a risk, our priority policy does not prioritise all the flits of a data block; rather, it just prioritises up to the critical flit. For example, if the critical flit is B1 for a data block, the proposed policy just prioritises B0 and B1 of that block. This is done because all the flits preceding the critical flit of the block should reach L1 at the earliest, then only the critical flit will reach. Once the critical flit is prioritised, the succeeding flits of the block (B2 and T) can reach L1 at their own pace as they are not required to resume processor execution.

\begin{figure*}
     \centering
     \begin{subfigure}[c]{0.496\textwidth}
         \centering
         \includegraphics[width=\textwidth, height=3cm]{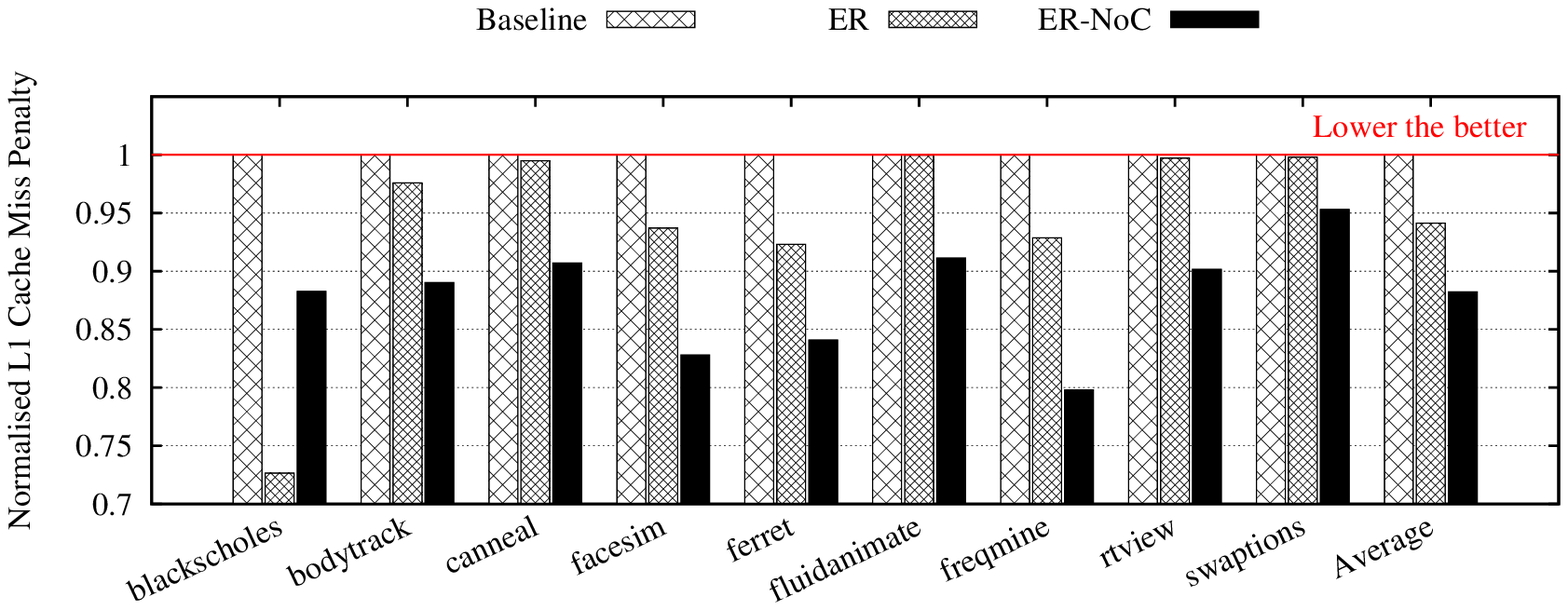}
         \caption{L1 Miss Penalty - PARSEC 3.0}
         \label{fig:misspen-p}
     \end{subfigure}
     \hfill
     \begin{subfigure}[c]{0.496\textwidth}
         \centering
         \includegraphics[width=\textwidth, height=3cm]{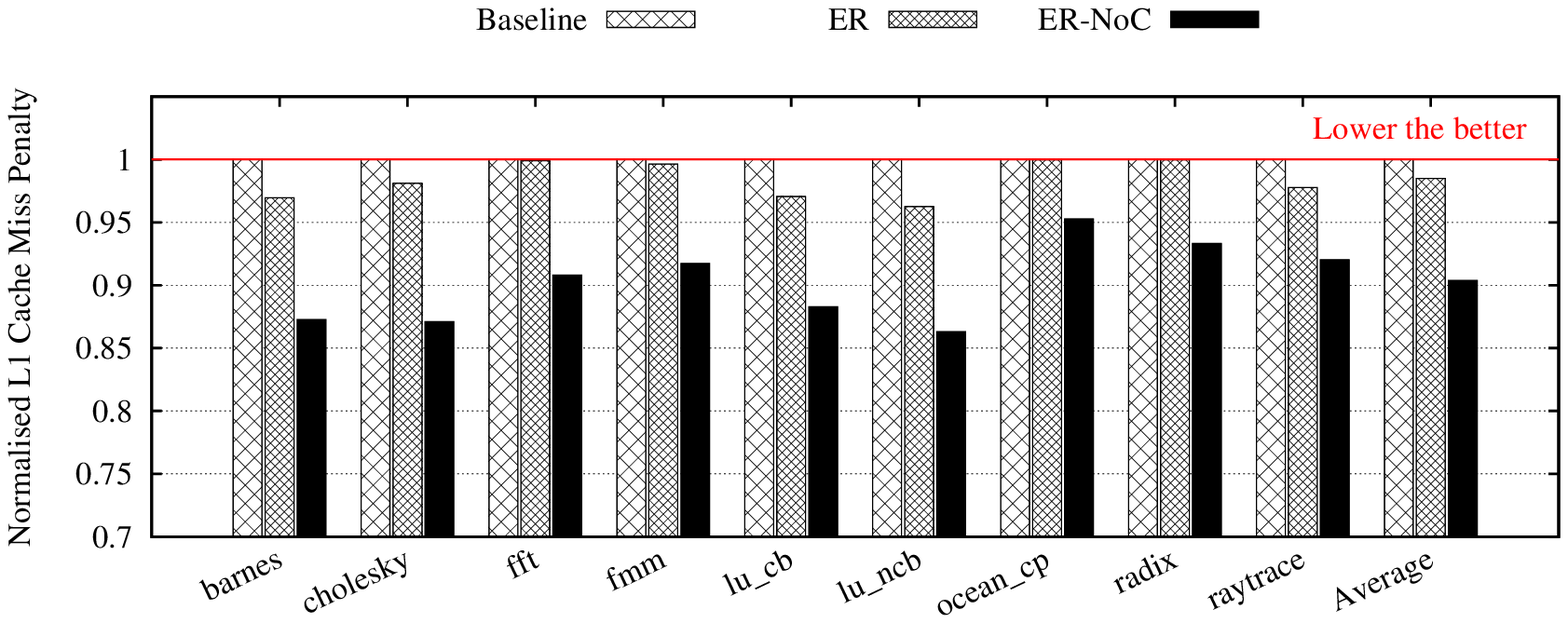}
         \caption{L1 Miss Penalty - SPLASH-2x}
         \label{fig:misspen-s}
     \end{subfigure}

     \begin{subfigure}[c]{0.496\textwidth}
         \centering
         \includegraphics[width=\textwidth, height=3cm]{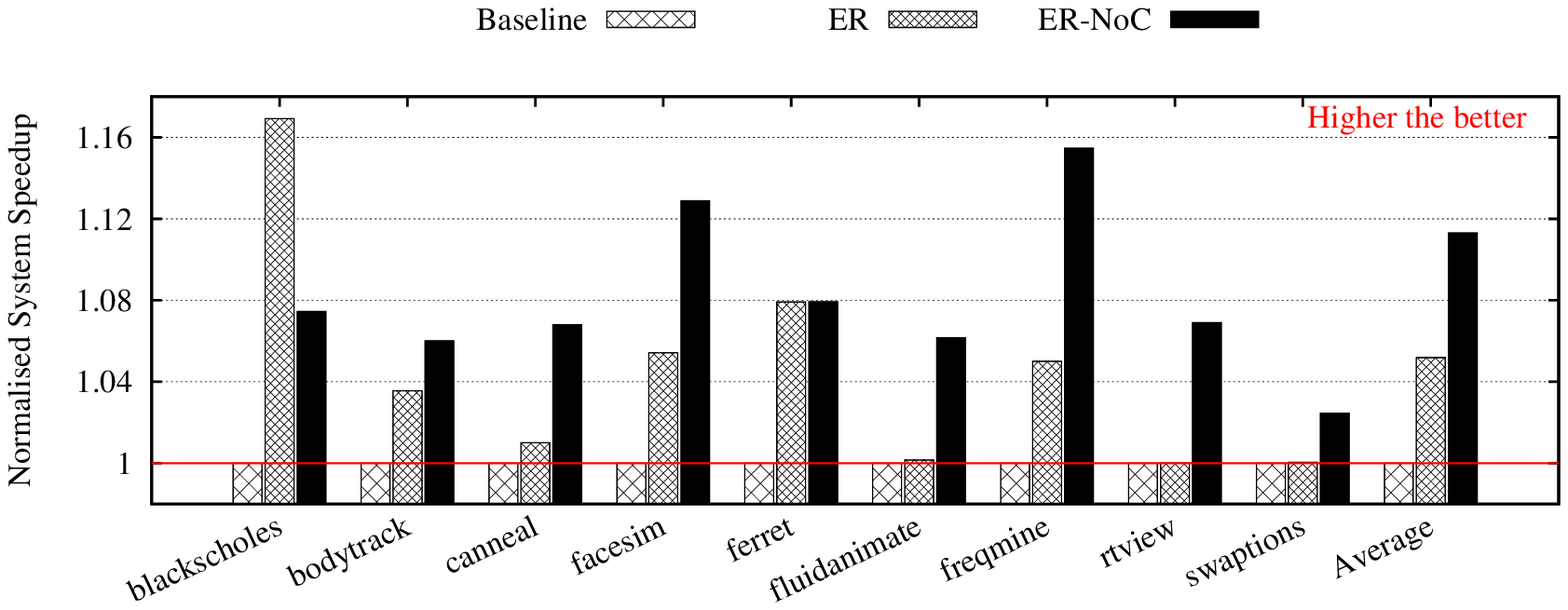}
         \caption{System Speedup - PARSEC 3.0}
         \label{fig:speed-p}
     \end{subfigure}
     \hfill
     \begin{subfigure}[c]{0.496\textwidth}
         \centering
         \includegraphics[width=\textwidth, height=3cm]{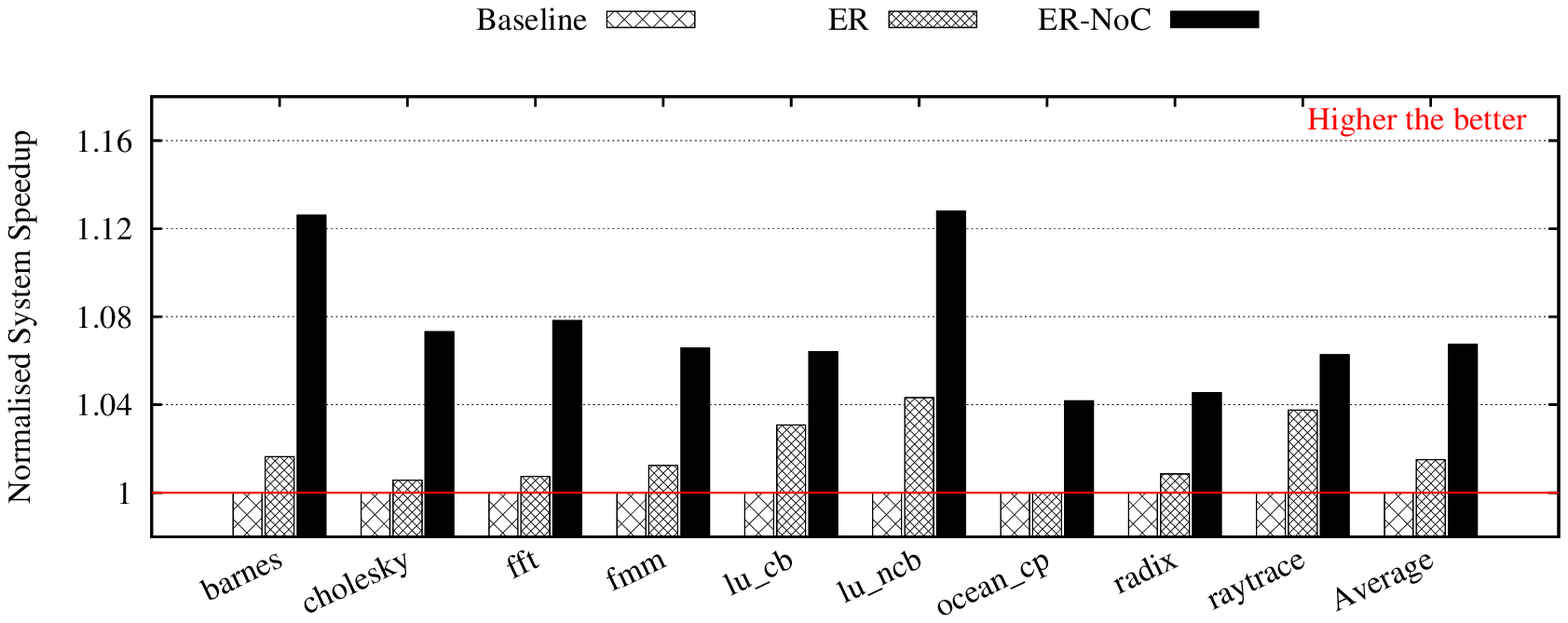}
         \caption{System Speedup - SPLASH-2x}
         \label{fig:speed-s}
     \end{subfigure}
     \caption{Performance of the proposed architecture}
     \label{fig:perf}
     \vspace{-0.6cm}
\end{figure*}

When two flits of two different block compete for the same output port, one with the lower CFI counter is prioritised. When a flit wins the arbitration, the corresponding CFI counter is decremented by 1. This way, when the critical flit reaches the router, the counter becomes 0, meaning the highest priority. Hence the proposed policy prioritises only and until critical flit of a data block. Exploiting NoC infrastructure to prioritise based on data criticality reduces miss penalty and improve overall system performance. This brief does not claim that the proposed priority policy gives the lowest starvation and best performance. The purpose is to show that it is beneficial to make existing ER like memory access optimisations aware of the NoC infrastructure in NoC-based many-core systems.

\subsection{Experimental Evaluation}
\noindent We consider the following architectures for evaluation:

\begin{itemize}
    \item \textbf{Baseline:} Without any optimisation.
    \item \textbf{ER:} Original {\it early restart}  optimisation.
    \item \textbf{ER-NoC:} Proposed NoC-aware {\it early restart} optimisation.
\end{itemize}

\noindent The system configuration used for evaluation is already given in Table~\ref{tab:config}. Each multi-threaded benchmark runs 64 threads in 64 different cores of the system. We use \textit{sim-medium}~\cite{bienia2008parsec} input set and report the results for the run of entire region-of-interest. We consider miss penalty and system speedup as the performance metrics for evaluation. All the results are normalised with respect to the baseline architecture.

\noindent \textbf{L1 Miss Penalty:} It is defined as the number of cycles required to replace an existing data block from L1 cache with the incoming requested block. Miss penalty directly reflects the effectiveness of the proposed ER-NoC optimisation. Figure~\ref{fig:misspen-p} and \ref{fig:misspen-s} shows the normalised miss penalty for PARSEC 3.0 and SPLASH-2x benchmarks suites. The existing ER optimisation reduces miss penalty by 6\% for PARSEC 3.0 and 2\% for SPLASH-2x. Exploiting the observed insights, our proposed ER-NoC performs better and significantly reduces miss penalty by 12\% and 10\%, respectively. Our prioritisation scheme is not optimal as we can see it performs poorly for \textit{blackscholes} when compared to ER. The focus is not to propose an optimal prioritisation scheme, rather highlighting the optimisation opportunities by exploiting the insights.

\noindent \textbf{System Speedup:} System speedup (S) is given by, $S = \frac{ExecTime_{baseline}}{ExecTime_{proposed}}$, where $ExecTime_{baseline}$ and $ExecTime_{proposed}$ are the execution time of baseline and proposed architectures, respectively. Figure~\ref{fig:speed-p} and \ref{fig:speed-s} show the normalised system speedup for PARSEC 3.0 and SPLASH-2x benchmarks suites. As expected, reduction in miss penalty directly directly translates to the improvement of overall system performance. Our proposed ER-NoC optimisation achieves an average system speedup of 11\% and 7\% for PARSEC 3.0 and SPLASH-2x suites, respectively.

\section{Related Works}
The existence of critical word and memory access optimisations to prioritise critical words are available on the classic textbook by Hennessy and Patterson~\cite{hennessy2011computer}. One of the first attempts to understand criticality for data requests from L1 to L2 cache is by Gieske~\cite{gieske2008critical}. He reported that for multi-programmed benchmark suites SPEC CPU 2000 and 2006, sufficient regularity in critical word exists. Exploiting this criticality information, quite a few optimisations are proposed in the recent past~\cite{chatterjee2012leveraging}\cite{lilly2014critical}\cite{huang2014increasing}\cite{li2016runahead}. However, none of them explicitly studied the behaviour of critical words for multi-threaded benchmark suites like PARSEC 3.0 and SPLASH-2x.

\section{Conclusion and Future Work}
In this brief, we shared two crucial insights about running multi-threaded applications in NoC based many-core systems. First, we show that critical words follow a particular pattern when requested from the memory. Then, we show that flits carrying the critical words gets indefinitely delayed in NoC. Finally, taking an existing memory access optimisation as a case study, we demonstrated that an NoC-aware optimisation can effectively utilise the two observed insights to significantly improve the overall system performance. In our future work, we will explore the utilisation of the insights to optimise existing memory access optimisations and apply them for developing an optimal critical word prioritisation policy.

% references section
\bibliographystyle{IEEEtran}
\bibliography{reference}

\end{document}